\documentclass{article}
\usepackage{itw2004}  
\usepackage{epsfig,psfig,rotating,amsmath,amssymb}

\newtheorem{thm}{Theorem}
\newtheorem{coro}{Corollary}
\newtheorem{lemma}{Lemma}
\newtheorem{example}{Example}

\newcommand{\newcaption}[1]{\caption{\footnotesize{#1}}}

\newcommand{\exit}{{\sf EXIT}}
\newcommand{\gexit}{{\sf GEXIT}}
\newcommand{\mmse}{{\sf MMSE}}

\def\cX{{\cal X}}
\def\cY{{\cal Y}}
\def\uX{\underline{X}}
\def\uY{\underline{Y}}
\def\0t{{\tt 0}}
\def\1t{{\tt 1}}
\def\sfa{{\sf a}}
\def\sfb{{\sf b}}
\def\tk{\tilde{k}}
\def\Ct{{\tt C}}
\def\snr{{\sf snr}}
\def\R{{\mathbb R}}
\def\E{{\mathbb E}}

\begin{document}
\topmargin = 0mm

\itwtitle{Life Above Threshold:\\ From List Decoding to Area Theorem
 and MSE\footnote{}}




 \itwauthor{Cyril M\'easson and R\"udiger Urbanke}
           {EPFL, I\&C \\
            CH-1015 Lausanne, Switzerland\\
            e-mail: {\tt cyril.measson@epfl.ch}\\ {\tt ruediger.urbanke@epfl.ch}
           }


  \itwsecondauthor{Andrea Montanari}
                 {
		   LPTENS (UMR 8549, CNRS et  ENS) \\
		   24, rue Lhomond, 75231  \\
		   Paris CEDEX 05, France \\
                   e-mail: {\tt montanar@lpt.ens.fr }
                 }

  \itwthirdauthor{Tom Richardson}
                  {
		    Flarion Technologies\\
		    Bedminster, NJ,  USA-07921\\
                    e-mail: {\tt richardson@flarion.com}
                  }




\itwmaketitle

\footnotetext[1]{{\tiny Copyright 2004 IEEE. Published in the 2004 IEEE Information Theory Workshop (ITW 2004), scheduled for October 24-29, 2004 at the Riverwalk Marriott in San Antonio, Texas, USA. Personal use of this material is permitted. However, permission to reprint/republish this material for advertising or promotional purposes or for creating new collective works for resale or redistribution to servers or lists, or to reuse any copyrighted component of this work in other works, must be obtained from the IEEE. Contact: Manager, Copyrights and Permissions / IEEE Service Center / 445 Hoes Lane / P.O. Box 1331 / Piscataway, NJ 08855-1331, USA. Telephone: + Intl. 732-562-3966.}}

\begin{itwabstract}
We consider communication over memoryless channels using low-density 
parity-check code ensembles above the iterative (belief propagation) 
threshold. What is the computational complexity of decoding (i.e., of
reconstructing {\it all}\, the typical input codewords for a given channel
output) in this regime? We define an algorithm accomplishing this task 
and analyze its typical performance. The behavior of the new algorithm
can be expressed in purely information-theoretical terms. 
Its analysis provides an alternative proof of the area theorem 
for the binary erasure channel. Finally, we explain how the area 
theorem is generalized to arbitrary memoryless channels. 
We note that the recently discovered relation
between mutual information and minimal square error is
an instance of the area theorem in the setting of Gaussian channels.
\end{itwabstract}

\begin{itwpaper}

\itwsection{Introduction}
    
The analysis of iterative coding systems has been extremely effective
in determining the conditions for successful communication. 
The single most important prediction in this context is the existence of a 
threshold noise level below which the bit error rate vanishes
(as the blocklength and the number of iterations diverge).
The threshold can be computed for a large variety of code ensembles 
using density evolution.

On the other hand, understanding the behavior of these systems 
{\em above threshold} is largely an open issue. Since in this regime the bit 
error rate remains bounded away from zero, one may wonder about
the motivation for such an investigation. We can think of three possible
answers: $(i)$ It is intellectually frustrating to have an ``half-complete''
theory of iterative decoding. Moreover this theory has poor connections with
classical issues such as the behavior of the same codes under
maximum likelihood (ML) decoding. $(ii)$ Loopy belief propagation 
has stimulated a considerable interest as a general-purpose inference 
algorithm for graphical models. However, there are very few 
applications where its effectiveness can be analyzed mathematically.
Decoding below threshold is probably the most prominent of such examples
and one may hope to build upon this success. $(iii)$ There are communication
contexts in which one is is interested in reproducing some information
within a pre-established tolerance, rather than exactly. 
There are indications that iterative methods can play an important
role also in such contexts. If this is the case, one will necessarily
operate in the above-threshold regime.

Consider, for the sake of simplicity, communication over a memoryless channel
using random elements from a standard low-density parity-check (LDPC) 
code ensemble. Assume moreover that the noise level is greater than the 
threshold one.
There are two natural theoretical problems one can address in this regime:
(A) How many channel inputs  correspond to a given typical output?
(B) How hard is to reconstruct {\em all} of them? 
Answering  question (A)
amounts to computing the conditional entropy  $H(X_1^n|Y_1^n)$ of 
the channel input given the output (here $n$ is the blocklength). 
We expect this entropy to become  of order $O(n)$ at large enough noise.
We call the minimum noise level for this to be the case, the ML 
threshold. ML decoding is bound to fail above this threshold.

The second question is apparently far from Information Theory and
in any case very difficult to answer.  
The naive expectation would be that reconstructing all the typical codewords
becomes harder as their  conditional entropy gets larger.

In this paper we report some recent progress on both of the questions outlined
above. In Secs.~\ref{sec:AreaBEC} and \ref{sec:Maxwell} we reconsider 
the binary erasure channel (BEC). We define a natural extension of 
the belief propagation decoder which reconstruct all the codewords
compatible with a given channel output. The new algorithm 
(`Maxwell decoder') thus performs a `complete' list decoding, 
and is based on the general message-passing
philosophy. Below the iterative threshold, it coincides with belief 
propagation decoding and its complexity is linear in the blocklength.
Above the iterative threshold, its complexity becomes exponential.
Its behavior can be analyzed precisely, and provides answers
both questions (A) and (B) above (within this circumscribed context).
Surprisingly, the resulting picture is most easily conveyed 
using a well-known information theoretic
characterization of the code: the $\exit$ curve. As a byproduct,
we obtain an alternative proof of the area theorem for the BEC.

The connection between the $\exit$ curve and Maxwell decoder is not
a peculiarity of the binary erasure channel, and has instead a 
rather fundamental origin.
The algorithm progressively reduces the uncertainty on the 
transmitted bits. This can be regarded as an effective change of the noise 
level of the communication channel. The $\exit$ curve describe the response 
of the bits (i.e., the change of the bit uncertainty) to a change 
in the noise level.
The area theorem is obtained when integrating this response: the total
bit uncertainty at maximal noise level (the code rate) is thus given
by an integral of the $\exit$ curve.

In Sec.~\ref{sec:General}, we explain how to generalize these
ideas to arbitrary memoryless channels. In particular, we define a
generalized $\exit$ function $\gexit$, which has the same important properties
of the usual one. We show that an area theorem holds for such a function,
implying, among other things, an upper bound on the ML threshold. 
$\gexit$  reduces to $\exit$ for the BEC and to the minimal mean-square
error ($\mmse$) for additive Gaussian channels.\vspace{1cm}\\
%
%
\itwsection{Area Theorem for the Binary Erasure Channel}
\label{sec:AreaBEC}

Consider a degree distribution pair $(\lambda, \rho)$ and ensembles 
${\rm LDPC}(n,\lambda,\rho)$ of increasing length $n$.
Figure \ref{fig:exitcurve} shows a typical asymptotic
$\exit$\footnote{The $\exit$ function is the function
$\frac{1}{n} \sum_{i=1}^{n} H(X_i|Y_{[n] \setminus \{ i\}})$, see 
\cite{Ten01}.} function.
Its main characteristics (for a regular ensemble with left degree at least 3) 
are as follows:
The function is zero below the ML threshold $\epsilon_{\rm{ML}}$. 
It jumps at $\epsilon_{\rm{ML}}$ to a non-zero value and
continues then smoothly until it reaches one for $\epsilon=1$. The area under
the $\exit$ curve equals the rate of the code, see \cite{AKtB02}. Compare 
this to the equivalent function of the iterative (IT) 
decoder which is also shown in Fig.~\ref{fig:exitcurve}.
It is easy to check that this curve is given in parametric form by
\begin{equation} \label{equ:exitcurve}
\left(
\frac{x}{\lambda(1-\rho(1-x))}, \Lambda(1-\rho(1-x))
\right),
\end{equation}
where $x$ signifies the erasure probability of left-to-right messages. 
Equation (\ref{equ:exitcurve}) can be derived from the fixed-point 
equation $\epsilon \lambda(1-\rho(1-x))-x=0$. We express
$\epsilon$ as $\epsilon(x)=\frac{x}{\lambda(1-\rho(1-x))}$ and 
 notice that the average probability
that a bit is still erased (ignoring the observation of the bit itself) 
at the fixed point is equal to $\Lambda(1-\rho(1-x))$.
Note that the iterative curve is the trace of this parametric equation for $x$
starting at
$x=1$ until $x=x_{\rm{IT}}$. This is the critical point  
and $\epsilon(x_{\rm{IT}})=\epsilon_{\rm{IT}}$.
Summarizing, the iterative $\exit$ curve is zero up to the iterative threshold 
$\epsilon_{\rm{IT}}$.
It then jumps to a non-zero value and also continues smoothly until it reaches
one at $\epsilon=1$. Multiple jumps are possible in some
irregular ensembles, but we shall neglect this possibility here.

The following two curious relationships between these two
curves were shown in \cite{MU03b}. 
First, the IT and the ML curve coincide above $\epsilon_{\rm{ML}}$.
Second, the ML curve can be constructed from the iterative curve in the 
following way.
If we draw the IT curve as parametrized in Eq.~(\ref{equ:exitcurve}) not only
for $x \geq x_{\rm{IT}}$ but also 
for $0 \leq x \leq x_{\rm{IT}}$ we get the
curve shown in the right picture of Fig.~\ref{fig:exitcurve}. 
Notice that the branch  $0 \leq x \leq x_{\rm{IT}}$ describes an unstable
fixed point under iterative decoding. Moreover the fraction of erased 
messages $x$ decreases along this branch when the erasure probability 
is increased. Finally it satisfies $x>\epsilon$.
Because of these  peculiar features, it is usually
considered as ``spurious''.

To determine the ML threshold take a straight vertical line at 
$\epsilon=\epsilon_{\rm{IT}}$ and shift it  to the right until the area 
which lies to the left of this straight line and is enclosed by the line
and the iterative curve is equal to the area which lies to the right 
of the line and is enclosed by the line and the iterative curve 
(these areas are indicated in dark gray in
the picture).
This unique point determines the ML threshold. 
The ML $\exit$ curve is now the curve which is zero to the left of the 
threshold and equals the iterative curve to the right
of this threshold. In other words, the ML threshold is determined by a balance
between two areas\footnote{The ML threshold was first determine
by the replica method in \cite{Mon01}.
Further, in \cite{Mon01b} a simple counting argument leading to an upper bound
for this threshold was given. In this paper we take as a starting point the 
point of view taken in \cite{MU03b}.}.
\begin{figure}[htp]
\begin{center}
\setlength{\columnsep}{0.5cm}
\setlength{\columnseprule}{0.0pt}
\begin{tabular}{cc}
\setlength{\unitlength}{1bp}
\begin{picture}(0,120)
\psfig{file=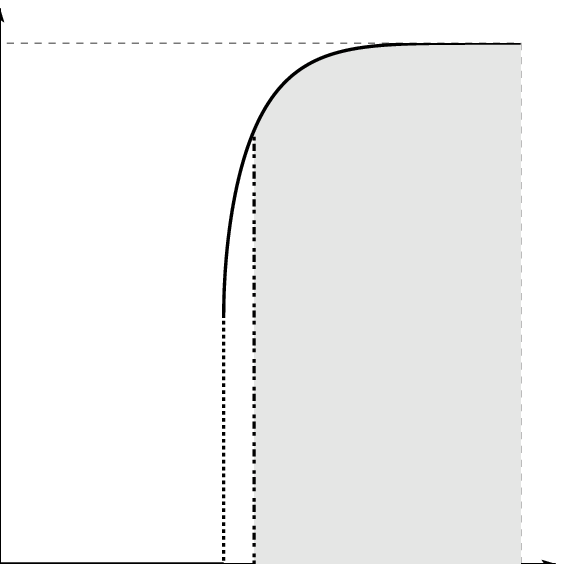,scale=0.6}
\end{picture}%
\begin{picture}(100,100)
\put(-5,-5){\makebox(0,0){$0$}}
\put(90,-5){\makebox(0,0){$1$}}
\put(-5,90){\makebox(0,0){$1$}}
\put(37, -5){\makebox(0,0){$\epsilon_{\rm IT}$}}
\put(53, -5){\makebox(0,0){$\epsilon_{\rm ML}$}}
\put(22, 80){\makebox(0,0){$\exit(\epsilon)$}}
\end{picture}
&
\begin{picture}(0,120)
\psfig{file=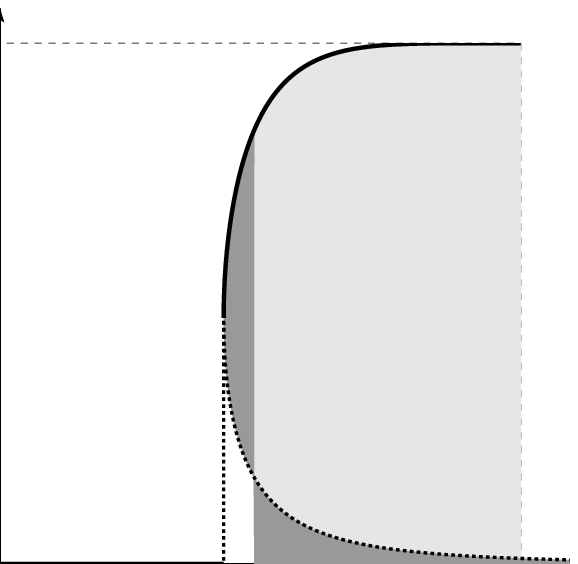,scale=0.6}
\end{picture}%
\begin{picture}(100,100)
\put(-3,-5){\makebox(0,0){$0$}}
\put(92,-5){\makebox(0,0){$1$}}
\put(-3,90){\makebox(0,0){$1$}}
\put(37, -5){\makebox(0,0){$\epsilon_{\rm IT}$}}
\put(53, -5){\makebox(0,0){$\epsilon_{\rm ML}$}}
\put(22, 80){\makebox(0,0){$\exit(\epsilon)$}}
\end{picture}
\end{tabular}
\newcaption{\label{fig:exitcurve}
Left: The $\exit$ curve of the ML decoder for the degree distribution pair 
$(\lambda(x)=x^2, \rho(x)=x^5)$.
The curve is zero until $\epsilon^{\rm{ML}}$ at which point it jumps. It then
 continuous smoothly
until it reaches one at $\epsilon=1$. Also shown is the equivalent curve under
iterative decoding. Right: The full iterative $\exit$ curve including 
the ``spurious branch''.  This corresponds to an unstable fixed point 
$x > \epsilon$. The ML threshold is determined by the balance of the two dark 
gray areas.}
\end{center}
\end{figure}
%
%
\itwsection{Maxwell Decoder}
\label{sec:Maxwell}

The balance condition described above, cf. Fig.~\ref{fig:exitcurve},
is strongly reminiscent of the so-called `Maxwell construction' in
statistical mechanics~\cite{KK02}. This allows, for instance, to determine 
the location of a liquid-gas phase  transition, by balancing two areas 
in the pressure-volume phase diagram. The Maxwell construction is 
derived by considering a reversible transformation between the liquid and 
vapor phases. The balance condition follows from the observation that the
net work exchange along such a transformation must vanish at the phase 
transition point. 

Inspired by the statistical mechanics analogy, we shall explain the
balance condition determining the ML threshold by analyzing 
an algorithm which moves from the non zero-entropy branch 
to the zero-entropy branch of the $\exit$ curve. 
To this end we construct a fictitious decoder, which for
obvious reasons we name the {\em Maxwell decoder}. 
Instead of explaining the balance between the areas as shown in 
Fig.~\ref{fig:exitcurve} we will explain the balance of the two areas shown 
in Fig.~\ref{fig:balance}. Note that these two areas differ from the
previous ones only by a common term so that the condition for balance stays 
unchanged.

\begin{figure}[htp]
\begin{center}
\setlength{\columnsep}{0.5cm}
\setlength{\columnseprule}{0.0pt}
\begin{tabular}{cc}
\setlength{\unitlength}{1bp}
\begin{picture}(0,0)
\psfig{file=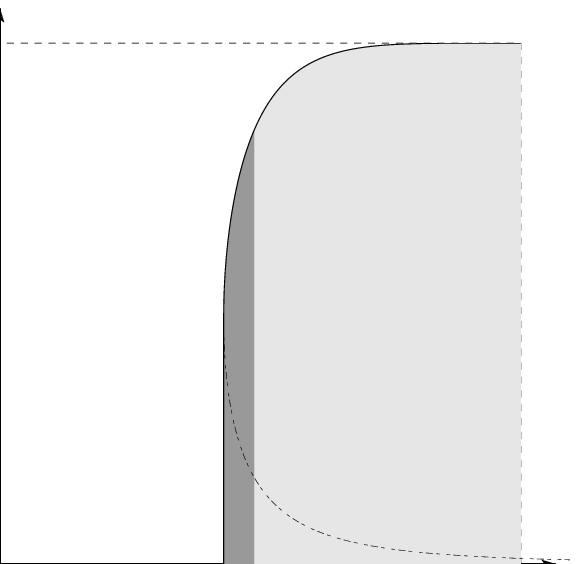,scale=0.6}
\end{picture}%
\begin{picture}(100,100)
\put(-5,-5){\makebox(0,0){$0$}}
\put(90,-5){\makebox(0,0){$1$}}
\put(-5,90){\makebox(0,0){$1$}}
\put(37, -5){\makebox(0,0){$\epsilon_{\rm IT}$}}
\put(53, -5){\makebox(0,0){$\epsilon_{\rm ML}$}}
\put(22, 80){\makebox(0,0){$\exit(\epsilon)$}}
\end{picture}
&
\begin{picture}(0,0)
\psfig{file=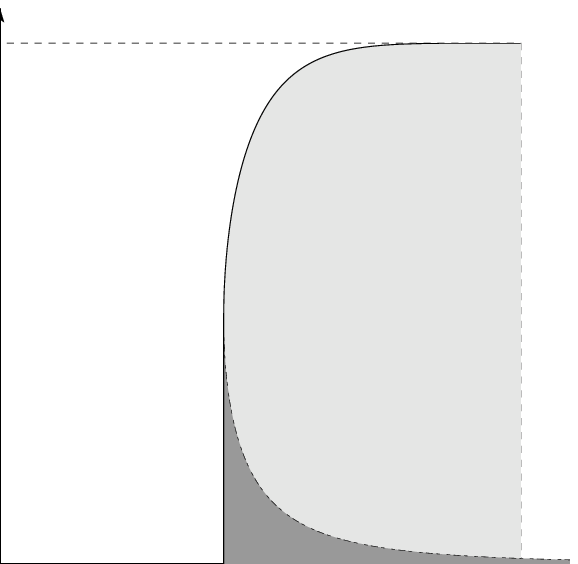,scale=0.6}
\end{picture}%
\begin{picture}(100,100)
\put(-3,-5){\makebox(0,0){$0$}}
\put(92,-5){\makebox(0,0){$1$}}
\put(-3,90){\makebox(0,0){$1$}}
\put(37, -5){\makebox(0,0){$\epsilon_{\rm IT}$}}
\put(53, -5){\makebox(0,0){$\epsilon_{\rm ML}$}}
\put(22, 80){\makebox(0,0){$\exit(\epsilon)$}}
\end{picture}
\end{tabular}
\newcaption{\label{fig:balance}$\exit$ curve for the $(x^2,x^5)$ ensemble. 
Dark gray: the two areas whose balance is proved by the analysis of the 
Maxwell decoder. Left: Total number of guesses made by a decoder starting at
$\epsilon_{\rm ML}$ (divided by the blocklength). Right: Total number of 
contradictions encountered (divided by the blocklength).}
\end{center}
\end{figure}
Let us now introduce the decoder: Given the received word which was
transmitted over the BEC$(\epsilon)$, the decoder proceeds iteratively as does
the standard message passing decoder. At any time the iterative
decoding process gets stuck in a non-empty stopping set the decoder randomly
chooses a position $i\in [n]$.  If this position is not known yet 
the decoder splits any running copy of the decoding process into two, 
one which proceed with the decoding process by assuming that $x_i=\0t$ 
and one which proceeds by assuming that $x_i=\1t$.
This splitting procedure is repeated any time the decoder gets stuck and 
we say that the decoder {\em guesses} a bit.
During the decoding it can
happen that contradictions occur, i.e., that a variable node
receives inconsistent messages. Any copy of the decoding process
which contains such contradictions terminates. From the above description
it follows that at any given point of the decoding process there are
$2^{h(\ell)}$ copies alive, where $h(\ell)$ is a natural number which 
evolves with time $\ell$.
Eventually, each surviving copies will has determined all the erased bits,
and outputs the corresponding word of size $n$. 
It is hopefully clear from the above description that the
final list of surviving copies is in one-to-one correspondence 
with the list of codewords that are compatible with the received message.
In other words, the Maxwell decoder performs a complete list decoding of
received message.

In Fig.~\ref{fig:maxwelltwodecoder} we depict an instance of the decoding 
process is shown from the perspective of the various simultaneous copies. 
The initial phase coincides with standard message passing:
a single copy of the process decodes a bit at a time. 
After three steps, belief propagation gets stuck in stopping 
set and  several steps of guessing follow. During this phase $h(\ell)$ 
(the associated entropy, i.e., the $\log_{2}$ of the
number of simultaneously running copies) increases. 
After this guessing phase, the standard message
passing phase resumes. More and more copies will terminate due to 
inconsistent messages (incorrect guesses).
At the end, only one copy survives, which shows that the example 
has a unique ML solution.\\

\begin{figure}[htp]
\begin{center}
\setlength{\unitlength}{0.8bp}
\newlength{\longueur}
\setlength{\longueur}{240bp}
\newcounter{Subdi}
\setcounter{Subdi}{20}
\newcounter{Step}
\setcounter{Step}{19}
\newcounter{Shift}
\setcounter{Shift}{0}
\begin{picture}(300,120)
\put(0,0){\psfig{file=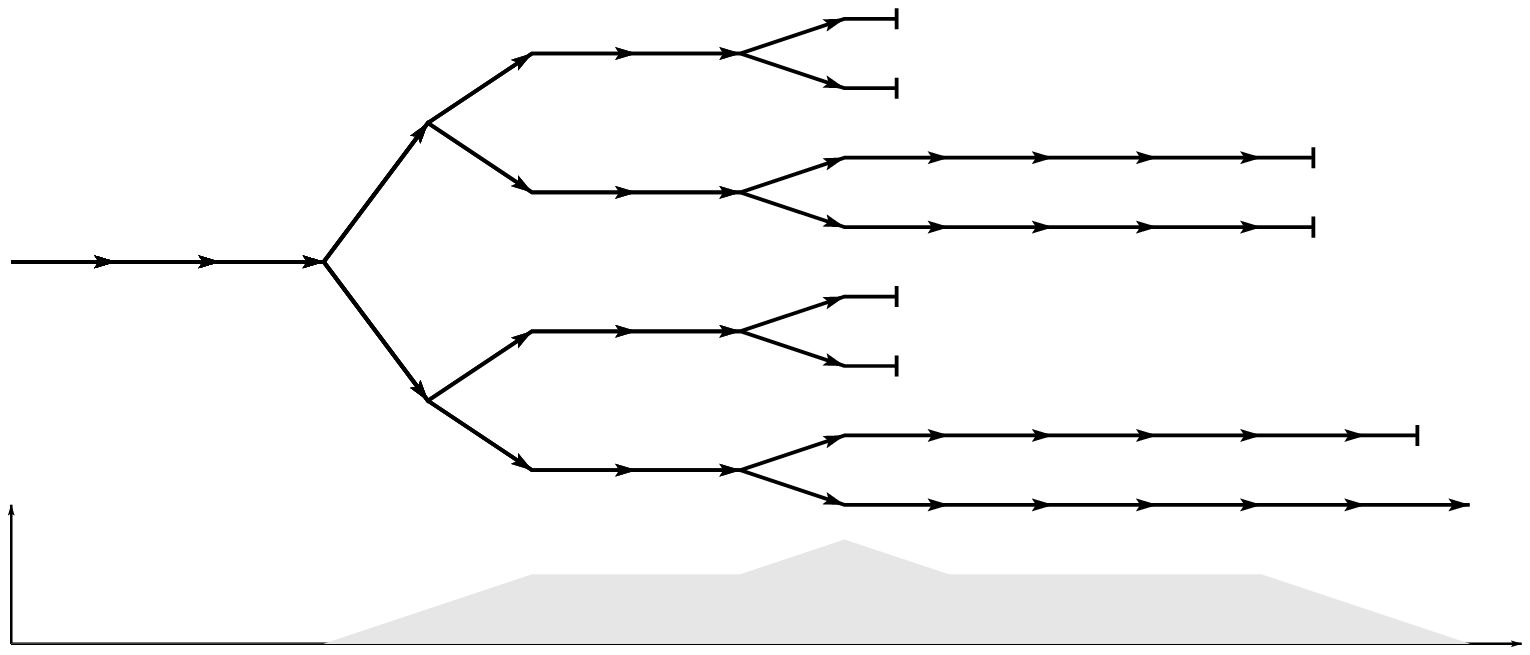,width=\longueur}}
\put(-3,38){\begin{turn}{0}\makebox(0,0){{\small $h(\ell)$} }\end{turn}}
\put(300,5){\begin{turn}{0}\makebox(0,0){{\small $\ell$} }\end{turn}}
\put(\value{Subdi},0){\begin{turn}{0}\makebox(0,0){{\small $x_1$} }\end{turn}}
\addtocounter{Subdi}{\value{Step}}
\put(\value{Subdi},0){\begin{turn}{0}\makebox(0,0){{\small $x_{10}$} }\end{turn}}
\addtocounter{Subdi}{\value{Step}}
\put(\value{Subdi},0){\begin{turn}{0}\makebox(0,0){{\small $x_{11}$} }\end{turn}}
\addtocounter{Subdi}{\value{Step}}
\put(\value{Subdi},0){\begin{turn}{0}\makebox(0,0){{\small $x_{2}$} }\end{turn}}
\addtocounter{Subdi}{\value{Step}}
\put(\value{Subdi},0){\begin{turn}{0}\makebox(0,0){{\small $x_{6}$} }\end{turn}}
\addtocounter{Subdi}{\value{Step}}
\put(\value{Subdi},0){\begin{turn}{0}\makebox(0,0){{\small $x_{28}$} }\end{turn}}
\addtocounter{Subdi}{\value{Step}}
\put(\value{Subdi},0){\begin{turn}{0}\makebox(0,0){{\small $x_{19}$} }\end{turn}}
\addtocounter{Subdi}{\value{Step}}
\put(\value{Subdi},0){\begin{turn}{0}\makebox(0,0){{\small $x_{12}$} }\end{turn}}
\addtocounter{Subdi}{\value{Step}}
\put(\value{Subdi},0){\begin{turn}{0}\makebox(0,0){{\small $x_{30}$} }\end{turn}}
\addtocounter{Subdi}{\value{Step}}
\put(\value{Subdi},0){\begin{turn}{0}\makebox(0,0){{\small $x_{24}$} }\end{turn}}
\addtocounter{Subdi}{\value{Step}}
\put(\value{Subdi},0){\begin{turn}{0}\makebox(0,0){{\small $x_{23}$} }\end{turn}}
\addtocounter{Subdi}{\value{Step}}
\put(\value{Subdi},0){\begin{turn}{0}\makebox(0,0){{\small $x_{21}$} }\end{turn}}
\addtocounter{Subdi}{\value{Step}}
\put(\value{Subdi},0){\begin{turn}{0}\makebox(0,0){{\small $x_{29}$} }\end{turn}}
\addtocounter{Subdi}{\value{Step}}
\put(\value{Subdi},0){\begin{turn}{0}\makebox(0,0){{\small $x_{26}$} }\end{turn}}
\addtocounter{Subdi}{\value{Step}}
\setcounter{Subdi}{20}
\addtocounter{Shift}{85}
\put(\value{Subdi},\value{Shift}){\begin{turn}{0}\makebox(0,0){{\small $0$} }\end{turn}}
\addtocounter{Subdi}{\value{Step}}
\put(\value{Subdi},\value{Shift}){\begin{turn}{0}\makebox(0,0){{\small $0$} }\end{turn}}
\addtocounter{Subdi}{\value{Step}}
\put(\value{Subdi},\value{Shift}){\begin{turn}{0}\makebox(0,0){{\small $0$} }\end{turn}}
\addtocounter{Subdi}{\value{Step}}
\addtocounter{Shift}{15}
\put(\value{Subdi},\value{Shift}){\begin{turn}{0}\makebox(0,0){{\small $1$} }\end{turn}}
\addtocounter{Shift}{-44}
\put(\value{Subdi},\value{Shift}){\begin{turn}{0}\makebox(0,0){{\small $0$} }\end{turn}}
\addtocounter{Subdi}{\value{Step}}
\addtocounter{Shift}{62}
\put(\value{Subdi},\value{Shift}){\begin{turn}{0}\makebox(0,0){{\small $1$} }\end{turn}}
\addtocounter{Shift}{-28}
\put(\value{Subdi},\value{Shift}){\begin{turn}{0}\makebox(0,0){{\small $0$} }\end{turn}}
\addtocounter{Shift}{-24}
\put(\value{Subdi},\value{Shift}){\begin{turn}{0}\makebox(0,0){{\small $1$} }\end{turn}}
\addtocounter{Shift}{-27}
\put(\value{Subdi},\value{Shift}){\begin{turn}{0}\makebox(0,0){{\small $0$} }\end{turn}}
\addtocounter{Subdi}{\value{Step}}
\addtocounter{Shift}{86}
\put(\value{Subdi},\value{Shift}){\begin{turn}{0}\makebox(0,0){{\small $0$} }\end{turn}}
\addtocounter{Shift}{-40}
\put(\value{Subdi},\value{Shift}){\begin{turn}{0}\makebox(0,0){{\small $1$} }\end{turn}}
\addtocounter{Shift}{-13}
\put(\value{Subdi},\value{Shift}){\begin{turn}{0}\makebox(0,0){{\small $1$} }\end{turn}}
\addtocounter{Shift}{-40}
\put(\value{Subdi},\value{Shift}){\begin{turn}{0}\makebox(0,0){{\small $0$} }\end{turn}}
\addtocounter{Subdi}{\value{Step}}
\addtocounter{Shift}{93}
\put(\value{Subdi},\value{Shift}){\begin{turn}{0}\makebox(0,0){{\small $1$} }\end{turn}}
\addtocounter{Shift}{-40}
\put(\value{Subdi},\value{Shift}){\begin{turn}{0}\makebox(0,0){{\small $0$} }\end{turn}}
\addtocounter{Shift}{-13}
\put(\value{Subdi},\value{Shift}){\begin{turn}{0}\makebox(0,0){{\small $1$} }\end{turn}}
\addtocounter{Shift}{-40}
\put(\value{Subdi},\value{Shift}){\begin{turn}{0}\makebox(0,0){{\small $0$} }\end{turn}}
\addtocounter{Subdi}{\value{Step}}
\addtocounter{Shift}{93}
\put(\value{Subdi},\value{Shift}){\begin{turn}{0}\makebox(0,0){{\small $1$} }\end{turn}}
\addtocounter{Shift}{-15}
\put(\value{Subdi},\value{Shift}){\begin{turn}{0}\makebox(0,0){{\small $0$} }\end{turn}}
\addtocounter{Shift}{-12}
\put(\value{Subdi},\value{Shift}){\begin{turn}{0}\makebox(0,0){{\small $1$} }\end{turn}}
\addtocounter{Shift}{-13}
\put(\value{Subdi},\value{Shift}){\begin{turn}{0}\makebox(0,0){{\small $0$} }\end{turn}}
\addtocounter{Shift}{-13}
\put(\value{Subdi},\value{Shift}){\begin{turn}{0}\makebox(0,0){{\small $1$} }\end{turn}}
\addtocounter{Shift}{-12}
\put(\value{Subdi},\value{Shift}){\begin{turn}{0}\makebox(0,0){{\small $0$} }\end{turn}}
\addtocounter{Shift}{-15}
\put(\value{Subdi},\value{Shift}){\begin{turn}{0}\makebox(0,0){{\small $1$} }\end{turn}}
\addtocounter{Shift}{-13}
\put(\value{Subdi},\value{Shift}){\begin{turn}{0}\makebox(0,0){{\small $0$} }\end{turn}}
\addtocounter{Subdi}{\value{Step}}
\addtocounter{Shift}{93}
\put(\value{Subdi},\value{Shift}){\begin{turn}{0}\makebox(0,0){{\small $~$} }\end{turn}}
\addtocounter{Shift}{-15}
\put(\value{Subdi},\value{Shift}){\begin{turn}{0}\makebox(0,0){{\small $~$} }\end{turn}}
\addtocounter{Shift}{-12}
\put(\value{Subdi},\value{Shift}){\begin{turn}{0}\makebox(0,0){{\small $1$} }\end{turn}}
\addtocounter{Shift}{-13}
\put(\value{Subdi},\value{Shift}){\begin{turn}{0}\makebox(0,0){{\small $0$} }\end{turn}}
\addtocounter{Shift}{-13}
\put(\value{Subdi},\value{Shift}){\begin{turn}{0}\makebox(0,0){{\small $~$} }\end{turn}}
\addtocounter{Shift}{-12}
\put(\value{Subdi},\value{Shift}){\begin{turn}{0}\makebox(0,0){{\small $~$} }\end{turn}}
\addtocounter{Shift}{-15}
\put(\value{Subdi},\value{Shift}){\begin{turn}{0}\makebox(0,0){{\small $1$} }\end{turn}}
\addtocounter{Shift}{-13}
\put(\value{Subdi},\value{Shift}){\begin{turn}{0}\makebox(0,0){{\small $0$} }\end{turn}}
\addtocounter{Shift}{93}
\addtocounter{Subdi}{\value{Step}}
\put(\value{Subdi},\value{Shift}){\begin{turn}{0}\makebox(0,0){{\small $~$} }\end{turn}}
\addtocounter{Shift}{-15}
\put(\value{Subdi},\value{Shift}){\begin{turn}{0}\makebox(0,0){{\small $~$} }\end{turn}}
\addtocounter{Shift}{-12}
\put(\value{Subdi},\value{Shift}){\begin{turn}{0}\makebox(0,0){{\small $1$} }\end{turn}}
\addtocounter{Shift}{-13}
\put(\value{Subdi},\value{Shift}){\begin{turn}{0}\makebox(0,0){{\small $0$} }\end{turn}}
\addtocounter{Shift}{-13}
\put(\value{Subdi},\value{Shift}){\begin{turn}{0}\makebox(0,0){{\small $~$} }\end{turn}}
\addtocounter{Shift}{-12}
\put(\value{Subdi},\value{Shift}){\begin{turn}{0}\makebox(0,0){{\small $~$} }\end{turn}}
\addtocounter{Shift}{-15}
\put(\value{Subdi},\value{Shift}){\begin{turn}{0}\makebox(0,0){{\small $1$} }\end{turn}}
\addtocounter{Shift}{-13}
\put(\value{Subdi},\value{Shift}){\begin{turn}{0}\makebox(0,0){{\small $0$} }\end{turn}}
\addtocounter{Subdi}{\value{Step}}
\addtocounter{Shift}{93}
\put(\value{Subdi},\value{Shift}){\begin{turn}{0}\makebox(0,0){{\small $~$} }\end{turn}}
\addtocounter{Shift}{-15}
\put(\value{Subdi},\value{Shift}){\begin{turn}{0}\makebox(0,0){{\small $~$} }\end{turn}}
\addtocounter{Shift}{-12}
\put(\value{Subdi},\value{Shift}){\begin{turn}{0}\makebox(0,0){{\small $0$} }\end{turn}}
\addtocounter{Shift}{-13}
\put(\value{Subdi},\value{Shift}){\begin{turn}{0}\makebox(0,0){{\small $1$} }\end{turn}}
\addtocounter{Shift}{-13}
\put(\value{Subdi},\value{Shift}){\begin{turn}{0}\makebox(0,0){{\small $~$} }\end{turn}}
\addtocounter{Shift}{-12}
\put(\value{Subdi},\value{Shift}){\begin{turn}{0}\makebox(0,0){{\small $~$} }\end{turn}}
\addtocounter{Shift}{-15}
\put(\value{Subdi},\value{Shift}){\begin{turn}{0}\makebox(0,0){{\small $1$} }\end{turn}}
\addtocounter{Shift}{-13}
\put(\value{Subdi},\value{Shift}){\begin{turn}{0}\makebox(0,0){{\small $0$} }\end{turn}}
\addtocounter{Subdi}{\value{Step}}
\addtocounter{Shift}{93}
\put(\value{Subdi},\value{Shift}){\begin{turn}{0}\makebox(0,0){{\small $~$} }\end{turn}}
\addtocounter{Shift}{-15}
\put(\value{Subdi},\value{Shift}){\begin{turn}{0}\makebox(0,0){{\small $~$} }\end{turn}}
\addtocounter{Shift}{-12}
\put(\value{Subdi},\value{Shift}){\begin{turn}{0}\makebox(0,0){{\small $0$} }\end{turn}}
\addtocounter{Shift}{-13}
\put(\value{Subdi},\value{Shift}){\begin{turn}{0}\makebox(0,0){{\small $1$} }\end{turn}}
\addtocounter{Shift}{-13}
\put(\value{Subdi},\value{Shift}){\begin{turn}{0}\makebox(0,0){{\small $~$} }\end{turn}}
\addtocounter{Shift}{-12}
\put(\value{Subdi},\value{Shift}){\begin{turn}{0}\makebox(0,0){{\small $~$} }\end{turn}}
\addtocounter{Shift}{-15}
\put(\value{Subdi},\value{Shift}){\begin{turn}{0}\makebox(0,0){{\small $1$} }\end{turn}}
\addtocounter{Shift}{-13}
\put(\value{Subdi},\value{Shift}){\begin{turn}{0}\makebox(0,0){{\small $0$} }\end{turn}}
\addtocounter{Shift}{93}
\addtocounter{Subdi}{\value{Step}}
\put(\value{Subdi},\value{Shift}){\begin{turn}{0}\makebox(0,0){{\small $~$} }\end{turn}}
\addtocounter{Shift}{-15}
\put(\value{Subdi},\value{Shift}){\begin{turn}{0}\makebox(0,0){{\small $~$} }\end{turn}}
\addtocounter{Shift}{-12}
\put(\value{Subdi},\value{Shift}){\begin{turn}{0}\makebox(0,0){{\small $~$} }\end{turn}}
\addtocounter{Shift}{-13}
\put(\value{Subdi},\value{Shift}){\begin{turn}{0}\makebox(0,0){{\small $~$} }\end{turn}}
\addtocounter{Shift}{-13}
\put(\value{Subdi},\value{Shift}){\begin{turn}{0}\makebox(0,0){{\small $~$} }\end{turn}}
\addtocounter{Shift}{-12}
\put(\value{Subdi},\value{Shift}){\begin{turn}{0}\makebox(0,0){{\small $~$} }\end{turn}}
\addtocounter{Shift}{-15}
\put(\value{Subdi},\value{Shift}){\begin{turn}{0}\makebox(0,0){{\small $1$} }\end{turn}}
\addtocounter{Shift}{-13}
\put(\value{Subdi},\value{Shift}){\begin{turn}{0}\makebox(0,0){{\small $0$} }\end{turn}}
\addtocounter{Shift}{93}
\addtocounter{Subdi}{\value{Step}}
\put(\value{Subdi},\value{Shift}){\begin{turn}{0}\makebox(0,0){{\small $~$} }\end{turn}}
\addtocounter{Shift}{-15}
\put(\value{Subdi},\value{Shift}){\begin{turn}{0}\makebox(0,0){{\small $~$} }\end{turn}}
\addtocounter{Shift}{-12}
\put(\value{Subdi},\value{Shift}){\begin{turn}{0}\makebox(0,0){{\small $~$} }\end{turn}}
\addtocounter{Shift}{-13}
\put(\value{Subdi},\value{Shift}){\begin{turn}{0}\makebox(0,0){{\small $~$} }\end{turn}}
\addtocounter{Shift}{-13}
\put(\value{Subdi},\value{Shift}){\begin{turn}{0}\makebox(0,0){{\small $~$} }\end{turn}}
\addtocounter{Shift}{-12}
\put(\value{Subdi},\value{Shift}){\begin{turn}{0}\makebox(0,0){{\small $~$} }\end{turn}}
\addtocounter{Shift}{-15}
\put(\value{Subdi},\value{Shift}){\begin{turn}{0}\makebox(0,0){{\small $~$} }\end{turn}}
\addtocounter{Shift}{-13}
\put(\value{Subdi},\value{Shift}){\begin{turn}{0}\makebox(0,0){{\small $0$} }\end{turn}}
\addtocounter{Subdi}{\value{Step}}
\end{picture}
\newcaption{\label{fig:maxwelltwodecoder} Maxwell decoder applied to a simple
example when the all-zero codeword is decoded.}
\end{center}
\end{figure}
In Fig.~\ref{fig:evol} we plot the  entropy $h(\ell)$ as a function
of the number of iterations for several code and channel realizations
(here we consider a $(3,6)$ ensemble with blocklength $n=10^4$ and erasure
probability $\epsilon=0.47$). It can be shown that the rescaled entropy
$h(\ell)/n$ concentrates around a finite limiting value if
we take the large blocklength limit $n\to\infty$, with $\ell/n$ fixed.
Moreover the limiting curve can be computed exactly. Here
we limit ourselves to outline the connection with
the various areas highlighted in Fig.~\ref{fig:balance}
and to explain why these areas should be in balance at the ML threshold.
To simplify matters consider only channel parameters $\epsilon$ with $\epsilon
\geq \epsilon_{\rm{IT}}$.
We claim that the total number of guesses one has to venture during the 
guessing phase of the algorithm is equal to the dark gray area 
shown in the left picture of Fig.~\ref{fig:balance}, 
i.e., it is equal to the integral under the iterative 
curve from $\epsilon_{\rm IT}$ up to $\epsilon$.

The effect of the guesses is to bring the effective erasure probability down 
from $\epsilon$ to $\epsilon_{\rm{IT}}$. At this point the standard message 
passing decoder can resume. 
The guesses are now resolved in the following manner. Assume that
at some point in time there is a variable node which has $d$ connected 
check nodes of degree one. The corresponding incoming messages have to be 
consistent. This gives rise to $d-1$ constraints, or in other words, 
only a fraction $2^{d-1}$ of the running copies survive. It can now 
be shown that the total number of such constraints which
are imposed is equal to the area in the left picture of Fig.~\ref{fig:balance}.
At the ML threshold all guesses have to be resolved at the end of the decoding
process. This implies that the total number of required guesses has 
to equal the total number of resolved guesses which 
implies an equality of the areas as promised!

\begin{figure}[htp]
\begin{center}
\epsfig{file=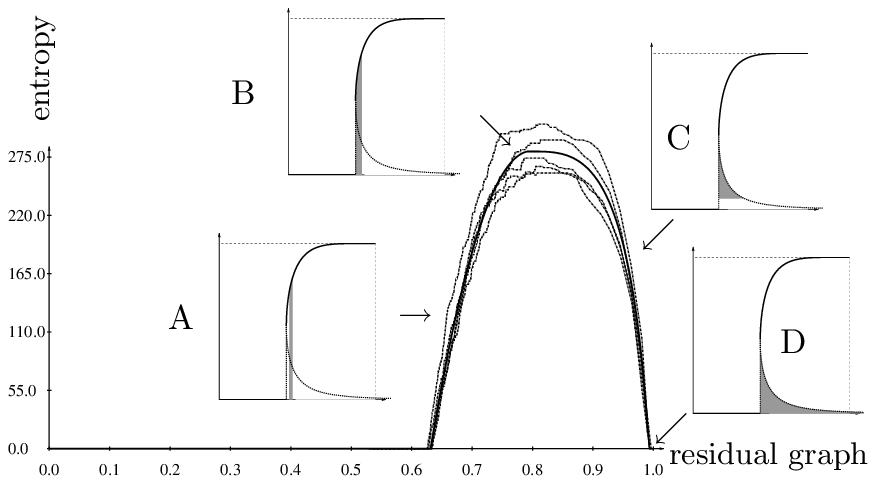,scale=0.8}
\newcaption{\label{fig:evol}Entropy of the Maxwell decoder 
(logarithm of the number of running copies) as a function of the
number of determined bits. We plot the results for several
channel and code realizations (here for a $(3,6)$ ensemble with 
blocklength $n=10^4$ and  $\epsilon=0.47$) together with the analytical 
asymptotic curve. In the inset: how the asymptotic curve can be constructed 
form the $\exit$ function.}
\end{center}
\end{figure}
Notice that the Maxwell decoder plays the same role as a reversible
transformation in thermodynamics.

%
%
\itwsection{General Channel}
\label{sec:General}

Three important lessons can be learned from the BEC example treated in the 
previous Sections. 
First of all: the $\exit$ curve gives the change in conditional 
entropy of the transmitted message when the channel noise level is incremented
by an infinitesimal amount. Second: in a search algorithm reconstructing 
all the typical input codewords, this change has to be compensated for by 
an increase of the algorithm entropy. This is the fundamental reason
of the equality between the area under the stable branch of the
$\exit$ curve and the number of guesses  made by the Maxwell decoder.
 Third: the fact that the iterative $\exit$ curve
extends below the maximum-likelihood one implies that the corresponding 
additional guesses must be eventually resolved. The unstable
branch of the $\exit$ curves yield the number of contradictions
found in this resolution stage.

The first step towards a generalization of this scenario for an arbitrary 
memoryless channel consist in finding the appropriate generalization 
of the $\exit$ curve. We obtain such a generalization by enforcing 
the first of the above properties. For the sake of definiteness,
we assume both the input and output alphabets to be finite and denote
by $Q(y|x)$, $x\in\cX$, $y\in\cY$ the transition probability. Formulae 
for continuous alphabets are easily obtained by substituting integrals
$\int\! dx$, $\int\! dy$, to sums $\sum_x$, $\sum_y$. We moreover denote 
by $w$ a generic noise-level parameter and assume  $Q(y|x)$ to be 
differentiable with respect to $w$. 
In analytical calculations, it is convenient to distinguish the noise levels
for each channel use $w_i$, $i\in [n]$. The time-invariant channel
is recovered by setting $w_1=\dots = w_n=w$   
Finally, we denote by $\uX\equiv X_1^n$
the channel input and $\uY\equiv Y_1^n$ the channel output. 
Our definition of a generalized $\exit$ curve is
\begin{eqnarray}
\gexit \equiv \frac{1}{n}\frac{d\phantom{w}}{dw}H(\uX|\uY)\, .
\end{eqnarray}
Notice that $\gexit$ satisfies the area theorem by construction:
our purpose is to get a manageable expression for it.
It is convenient to think of the above differentiation as acting on each 
channel separately
\begin{eqnarray}
\gexit = \frac{1}{n}\sum_{i=1}^n\frac{d\phantom{w}}{dw_i}H(\uX|\uY)
\equiv \frac{1}{n}\sum_{i=1}^n\gexit_i\, ,
\end{eqnarray}
with $\gexit_i$ defined as the derivative with respect to $w_i$.
In order to compute $\gexit_i$, it is convenient to isolate
the contribution of $X_i$ to the conditional entropy. If we denote 
by $Z_i$ the extrinsic information at $i$, and use the shorthands
$\uX^{[i]} \equiv(X_j\, :\, j\in[n]\backslash i)$,
$\uY^{[i]} \equiv(Y_j\, :\, j\in[n]\backslash i)$, we get
\begin{eqnarray}
H(\uX|\uY) = H(X_i | Z_i, Y_i) +  H( \uX^{[i]} | X_i, \uY^{[i]} )\, .
\label{eq:EntropyDecomposition}
\end{eqnarray}
This is obtained by a standard application of the entropy chain rule
\begin{eqnarray*}
H(\uX|\uY)  & = & H(X_i | \uY ) + H(\uX^{[i]}| X_i, \uY)\\
            & = & H(X_i | \uY,Z_i) + H(\uX^{[i]} | X_i, \uY^{[i]})\\
            & = & H(X_i | Y_i,\uY^{[i]},Z_i) +  H(\uX^{[i]} | X_i,\uY^{[i]})\\
            & = & H(X_i | Y_i,Z_i) +  H(\uX^{[i]} | X_i, \uY^{[i]})\, .
\end{eqnarray*}
We remark at this point that only the first term of the decomposition
(\ref{eq:EntropyDecomposition}) depends upon the channel at position $i$. 
Therefore
\begin{eqnarray}
\gexit_i = \frac{d\phantom{w}}{dw_i}H(X_i | Z_i, Y_i)\, .
\end{eqnarray}
It is convenient to obtain a more explicit expression for the above
formula. To this end we write
\begin{eqnarray}
H(X_i | Z_i, Y_i) &= &-\sum_{x_i,y_i,z_i}P(z_i)P(x_i|z_i)
Q(y_i|x_i)\cdot\\ 
&&\phantom{\sum_{x_i,y_i,z_i}}
\cdot\log\left\{\frac{P(x_i|z_i)Q(y_i|x_i)}{\sum_{x_i'\in\cX}
P(x'_i|z_i)Q(y_i|x'_i)}\right\}\, .\nonumber
\end{eqnarray}
The dependence of $H(X_i | Z_i, Y_i)$ upon the channel at position $i$ 
is completely explicit and we can differentiate. The terms obtained
by differentiating with respect to the channel {\em inside} the log
vanish. For instance, when differentiating with respect to the $Q(y_i|x_i)$
at the numerator, we get
\begin{eqnarray*}
 & &-\sum_{x_i,y_i,z_i}P(z_i)P(x_i|z_i)\frac{d\phantom{w}}{dw_i}
Q(y_i|x_i) \\ 
&=& -\sum_{x_i,z_i}P(z_i)P(x_i|z_i)\frac{d\phantom{w}}{dw_i}\sum_{y_i} Q(y_i|x_i) 
=0\, .
\end{eqnarray*}
We thus proved the following
\begin{thm}
With the above definitions
\begin{eqnarray}
\gexit_i &=&  \sum_{x_i,y_i,z_i}P(x_i)P(z_i|x_i)
Q'(y_i|x_i)\cdot\label{eq:GeneralEXIT}\\
&&\phantom{\sum_{x_i,y_i,z_i}}\cdot
\log\left\{\sum_{x'_i}\frac{P(x'_i|z_i)Q(y_i|x'_i)}{P(x_i|z_i)Q(y_i|x_i)}
\right\}\, ,\nonumber
\end{eqnarray}
where we denoted by $Q'(y|x)$ the derivative of the channel transition
probability with respect to the noise level $w$.
\end{thm}
The interest of the above result is that it encapsulates all our ignorance 
about the code behavior into the distribution of extrinsic information
$P(z_i)$. This is in turn the natural object appearing in message passing 
algorithms and in density evolution analysis. In order to fully
appreciate the meaning of Eq.~(\ref{eq:GeneralEXIT}), it is convenient
to consider a couple of more specific examples.

\underline{Linear Codes over BMS Channels}

We assume the code to be linear and to be used over a binary-input memoryless 
 output-symmetric (BMS) channel. We furthermore denote the
channel input by  $\cX = \{\0t,\1t\}$.
This is the most common setting in the analysis of iterative coding
systems. 
Exploiting the channel symmetry we can fix $x_i=\0t$ in 
Eq.~(\ref{eq:GeneralEXIT}) and get 
\begin{eqnarray}
\gexit_i = \sum_{y_i,z_i}P_0(z_i)
Q'(y_i|x_i)
\log\left\{1+\frac{P(\1t|z_i)Q(y_i|\1t)}{P(\0t|z_i)Q(y_i|\0t)}
\right\}\, ,\nonumber\\
\end{eqnarray}
where we defined $P_0(z_i)$ to be the distribution of the extrinsic
information at $i$ under the condition that the all-zero codeword has been 
transmitted. Recall that $Z_i$ is a function of $\uY^{[i]}$ and 
$P_0(z_i)$ is the distribution induced  on $Z_i$ by the distribution of
$\uY^{[i]}$.

It is convenient to encode the extrinsic information $z_i$ as an extrinsic 
log-likelihood ratio $l_i\equiv \log[P(\0t|z_i)/P(\1t|z_i)]$.
Analogously, we define $L_Q(y) \equiv \log[Q(y|\0t)/Q(y|\1t)]$.
Finally, we denote by $\sfa^{(i)}(l)$ the density of $l_i$ with respect to the
Lebesgue measure. We thus get the following handy expression
\begin{coro}
For a linear code over a BMS channel
\begin{eqnarray}
\gexit_i = \int_{-\infty}^{+\infty}\!\!\sfa^{(i)}(l) \; k_L(l)\; dl\, ,
\end{eqnarray}
where we introduced the $\gexit$ kernel
\begin{eqnarray}
k_L(l) \equiv \sum_{y} Q'(y|\0t)\log(1+e^{-L_Q(y)-l})\, .
\end{eqnarray}
\end{coro}
It is worth recalling that the usual $\exit$ curve has a similar expression.
In fact
\begin{eqnarray}
\exit_i = \int_{-\infty}^{+\infty}\!\!\sfa^{(i)}(l) \; \tk_L(l)\; dl\, ,
\end{eqnarray}
with the channel-independent $\exit$ kernel
$\tk_L(l) \equiv \log(1+e^{-l})$. Finally, we notice that it is possible 
to use alternative encodings for the extrinsic information.
One important possibility is to work in the so-called `difference
domain' $z =\tanh(l/2)$. The new kernel will be given by
$k_{D}(z) \equiv k_L(2\tanh^{-1}(z))$. It is moreover possible to exploit the 
symmetry property of $\sfa^{(i)}(l)$ to get
\begin{eqnarray}
\gexit_i = \int_{0}^{+\infty}\!\!\sfa^{(i)}(l) \; k_{|L|}(l)\; dl\, ,
\end{eqnarray}
where $k_{|L|}(l) = k_{L}(l) +e^{-l}k_{L}(l)$. Analogously, one can 
consider an `absolute difference' kernel $k_{|D|}(z)$.

Let us work out a couple of examples. 
In order to compare the different cases, it
is useful to define a unified convention for the noise level parameter
$w$. We choose $w$ to be the channel entropy, or, in other words,
one minus the channel capacity: $w=1-\Ct(Q)$ (in bits).

For the BEC we have $\cY = \{\0t,\1t,\ast\}$ and the transition probabilities
read $Q(\0t|\0t) = 1-\epsilon$, $Q(\ast|\0t) = \epsilon$, 
$Q(\1t|\0t) = 0$. Obviously $w=\epsilon$ and $Q'(\0t|\0t) = -1$,
$Q'(\ast|\0t) = 1$, $Q'(\1t|\0t) = 0$. We get
\begin{eqnarray}
k_L^{\rm BEC}(l) = \log(1+e^{-l})\, .
\end{eqnarray}
Therefore $k_L^{\rm BEC}(l) = \tk_L(l)$ and $\gexit_i = \exit_i$.
We thus recovered a well known result: the $\exit$ curve verifies  
the area theorem for the BEC.

\begin{figure}[hbt]
\begin{center}
\begin{picture}(360,90)
\put(0,0){\psfig{file=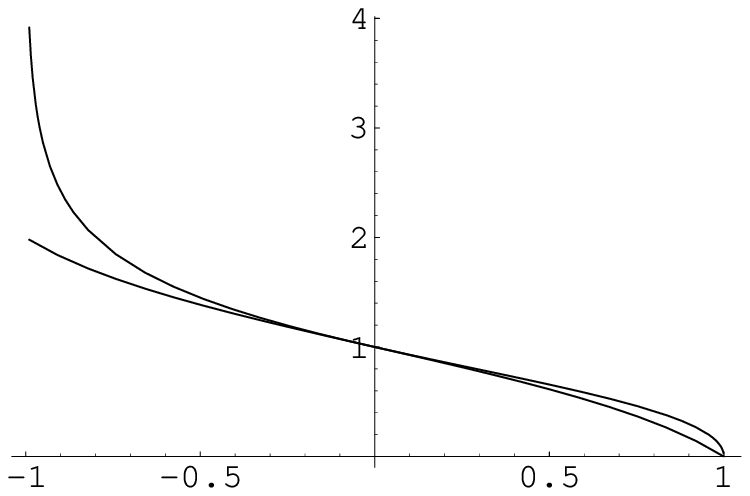,width=110bp}}
\put(120,0){\psfig{file=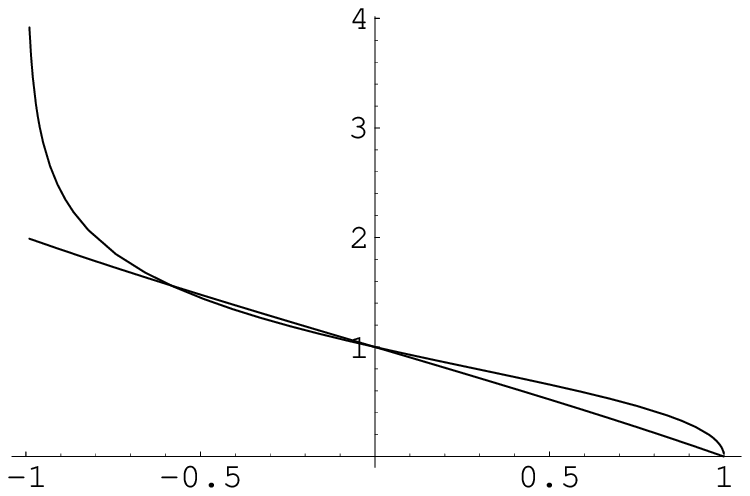,width=110bp}}
\end{picture}
\end{center}
 \newcaption{Difference-domain kernels $k^{\rm BSC}_{D}$ and $\tk^{\rm BSC}_{D}$ 
for $\gexit$ and $\exit$ curves on the BSC. They should be multiplied by the 
extrinsic information distribution in the
 $D$-domain
 and integrated over $(-1,1)$. 
Left: flip probability $p=0.1$. Right: $p=0.3$ }
 \label{fig:Dnk}
 \end{figure}
\begin{figure}[hbt]
\begin{center}
\begin{picture}(360,90)
\put(0,0){\psfig{file=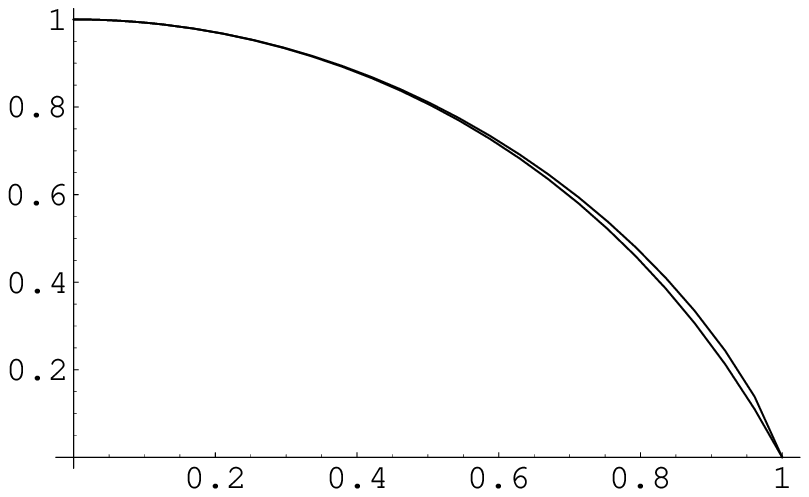,width=110bp}}
\put(120,0){\psfig{file=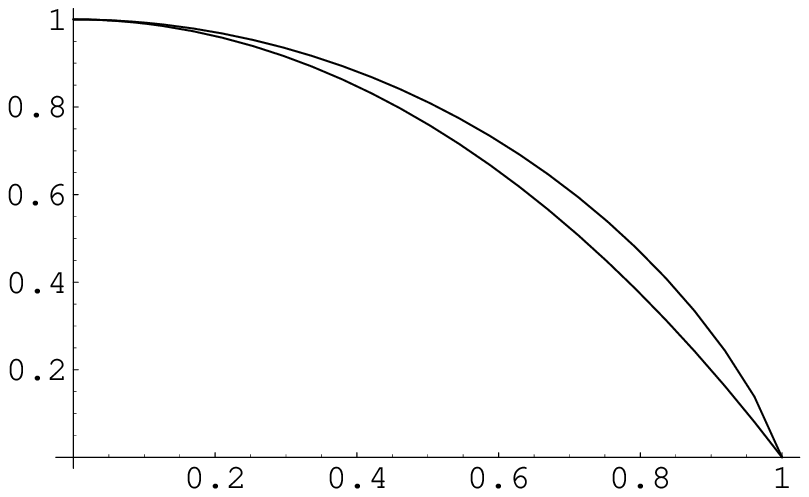,width=110bp}}
\end{picture}
\end{center}
 \newcaption{Absolute difference domain kernels $k^{\rm BSC}_{|D|}$ and 
$\tk^{\rm BSC}_{|D|}$ for $\gexit$ and $\exit$ curves on the BSC.
They should be multiplied by the extrinsic information distribution in the
 $D$-domain 
and integrated over  $[0,1)$. Left: flip probability $p=0.1$. Right: $p=0.3$ }
 \label{fig:Dk}
 \end{figure}
Consider now the BSC with flip probability $p$. Proceeding as above, we get
\begin{eqnarray}
k_L^{\rm BSC}(l) = \frac{1}{\log(\frac{1-p}{p})}\left[
\log(1+\frac{1-p}{p}\, e^{-l})-\log(1+\frac{p}{1-p}\, e^{-l})\right]\, ,
\nonumber
\hspace{-1.cm}\\
\end{eqnarray}
In Figs.~\ref{fig:Dnk} and \ref{fig:Dk} we plot the $\gexit$ kernel in the 
difference and absolute
difference domains, comparing it with the usual $\exit$ one.
In Fig.~\ref{fig:gexitbsc} we plot the $\exit$ and $\gexit$ curves for a 
few regular LDPC ensembles over the BSC.
\begin{figure}[hbt]
\begin{center}
\begin{picture}(200,200)
\put(0,0){\psfig{file=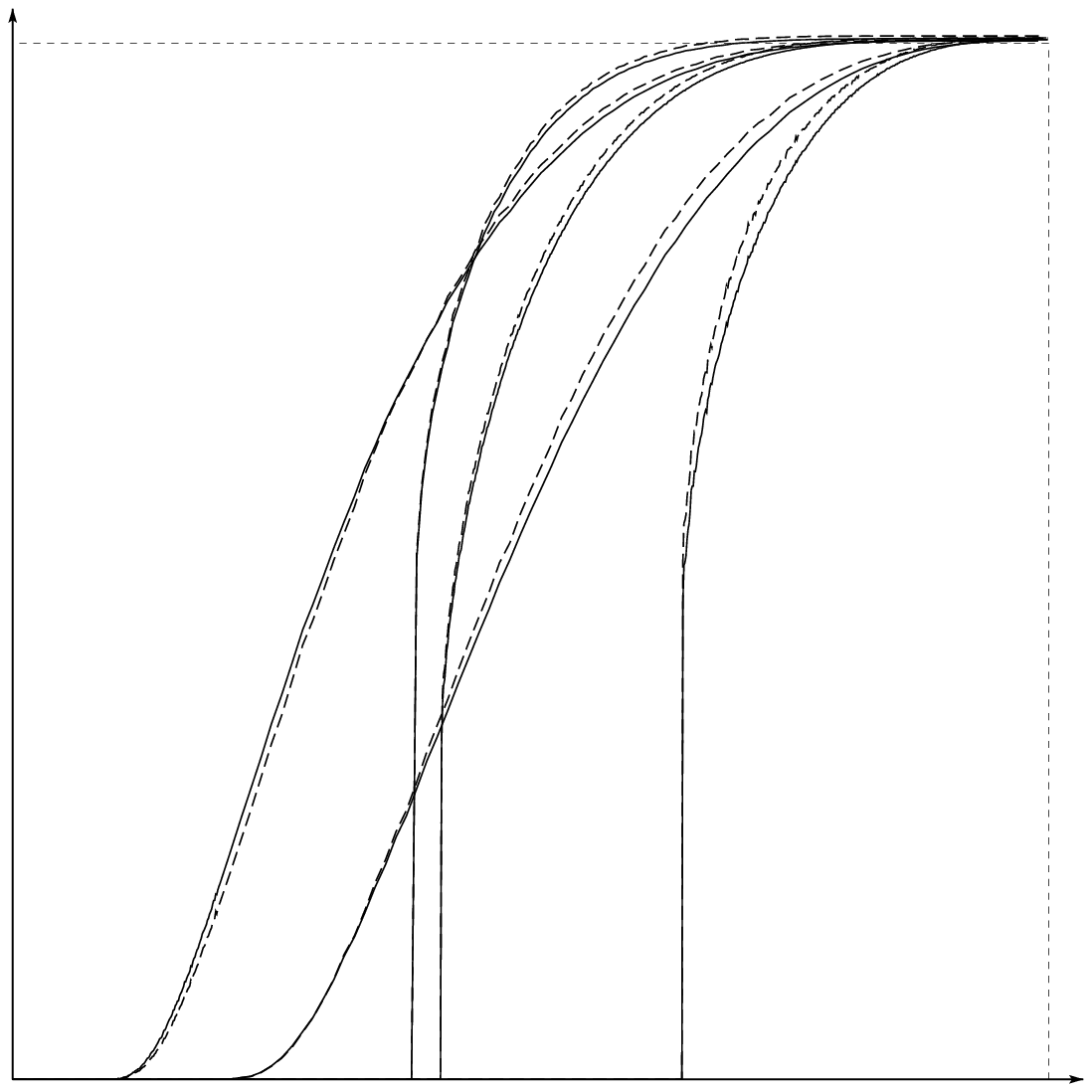,width=200bp}}
\put(130,30){$(3,4)$}
\put(85,30){$(3,6)$}
\put(57,70){$(4,8)$}
\put(48,40){$(2,4)$}
\put(16,40){$(2,6)$}
\put(0,-5){$0$}
\put(190,-5){$1$}
\put(210,-3){$p$}
\put(10,180){$\exit(p),\, \gexit(p)$}
\end{picture}
\end{center}
 \newcaption{$\gexit$ (solid curves) 
versus $\exit$ (dashed curves) for several 
regular LDPC ensembles over the BSC.}
 \label{fig:gexitbsc}
 \end{figure}

From these examples it should be clear that computing the $\gexit$ curve 
is not harder than computing the $\exit$ one. The  difference 
between these two curves is often quantitatively small (cf. for instance
Fig.~\ref{fig:gexitbsc}).  
Nevertheless such a difference is definitely different from
zero and it is not hard to show that an area theorem cannot hold
for the $\exit$ curve. 
Finally, several qualitative properties remain unchanged. In particular
\begin{lemma}
Given a density $\sfa(l)$ over the reals, let
\begin{eqnarray}
\gexit_{\rm BSC}(\sfa)\equiv \int_{-\infty}^{\infty}\!\!\sfa(l) \; 
k^{\rm BSC}_L(l)\; dl \, .
\end{eqnarray}
If the density $\sfb(l)$ is physically degraded with respect to
$\sfa(l)$, then $\gexit_{\rm BSC}(\sfb)\ge \gexit_{\rm BSC}(\sfa)$.
\end{lemma}
An important application of the above Lemma consists in approximating 
the correct extrinsic LLR densities with the site-averaged 
belief propagation density.  
This yields an upper bound on the $\gexit$ curve:
\begin{eqnarray*}
\gexit &= &\frac{1}{n}\sum_{i=1}^n \gexit_i \\ 
& = & \gexit_{\rm BSC}\left(\frac{1}{n}\sum_{i=1}^n \sfa^{(i)}\right)\\
&\le & \gexit_{\rm BSC}(\sfa^{{\rm BP},k})\, .
\end{eqnarray*}
where we denoted by $\sfa^{{\rm BP},k}$ the belief propagation density after 
$k$ iterations. We can now take the $n\to\infty$ limit and 
(afterwards) the $k\to\infty$ limit to get
\begin{eqnarray*}
\gexit \le \gexit_{\rm BSC}(\sfa^{{\rm DE},*})
\end{eqnarray*}
where $\sfa^{{\rm DE},*}$ is the density at the density evolution fixed point.
We obtain therefore the following
\begin{coro}
Consider communication over the BSC using LDPC$(n,\lambda,\rho)$ 
ensembles of rate $R$, and let $p_{\rm ML,\, DE}$ be defined by
\begin{eqnarray}
R = \int_{p_{\rm ML,\, DE}}^1 \gexit_{\rm BSC}(\sfa^{{\rm DE},*})\, dp
\, ,
\end{eqnarray}
with  $\sfa^{{\rm DE},*}$ the density at the density evolution fixed point
at flip probability $p$. Let moreover $p_{\rm ML}$ be the maximum likelihood 
threshold defined as the smallest noise level such that the ensemble-averaged 
conditional entropy $\E H(\uX|\uY)$ is linear in the blocklength. Then
\begin{eqnarray}
p_{\rm ML} \le p_{\rm ML,\, DE}\, .
\end{eqnarray}
\end{coro}
\begin{example}
For the $(3,6)$ ensemble and the BSC, the previous method gives $p_{\rm ML} \le 0.101$.
\end{example}

\underline{Gaussian Channels}

We assume $\cX=\cY = \R$ and 
\begin{eqnarray}
Q(y|x) = \frac{1}{\sqrt{2\pi}}\; \exp\left\{-\frac{1}{2}(y-\sqrt{\snr}\,
x)^2\right\}\, .
\end{eqnarray}
Notice that, in this case, $Q(y|x)$ should be interpreted as a density
with respect to Lebesgue measure. An alternative formulation of the same 
channel model consists in saying that $Y = \sqrt{\snr}X+W$ with 
$W$ a standard Gaussian variable.
It is also useful to define the minimal mean square error
$\mmse_i$ in estimating $X_i$ as follows
\begin{eqnarray}
\mmse_i \equiv \E_{Y_iZ_i}\left\{\E_{X_{i}}[x^2_i|y_i,z_i]-
\E_{X_{i}}[x_i|y_i,z_i]^2\right\}\, ,
\end{eqnarray}
where $\E_{X,Y,\dots}$ denotes expectation with respect to the variables
$\{X,Y,\dots\}$.
Finally, we take the signal-to-noise ratio
as the noise parameter entering in the definition of
the $\gexit$ curve: $w=\snr$. The reader will easily translate the
results to other choices of $w$ by a change of variable. 

As recently shown by Guo, Shamai and Verdu~\cite{Guo04},
the derivative with respect to the signal-to-noise ratio of the mutual
information across a gaussian channel is related to the minimal
mean-square error. Adapting their result to the present context,
we immediately obtain the following
\begin{coro}
For the additive Gaussian channel defined above, we have
\begin{eqnarray}
\gexit_i = - \frac{1}{2}\, \mmse_i\, .\label{eq:mmse}
\end{eqnarray}
\end{coro}
For greater convenience of the reader we briefly recall the 
derivation~\cite{Guo04}
of this result from the expression (\ref{eq:GeneralEXIT}).
In order to keep things simple,
we shall consider here the case of a simple symbol with input 
density $P(x)$ transmitted uncoded through the channel. The generalization
is immediate. 
We rewrite Eq.~(\ref{eq:GeneralEXIT}) in the single symbol case as
\begin{eqnarray}
\gexit = \int\!\!\!\int P(x) Q'(y|x)\,\log\left[\int\!
\frac{P(x')Q(y|x')}{P(x)Q(y|x)}
dx'\right] dx\; dy\, .\nonumber\\ \label{eq:shortentropy}
\end{eqnarray}
It is convenient to group at this point a couple of remarks which simplify
the calculations. First
\begin{eqnarray}
Q'(y|x) = -\frac{x}{2\sqrt{\snr}}\frac{d\phantom{y}}{dy}Q(y|x)\, .
\label{eq:twoderivatives}
\end{eqnarray}
Second
\begin{eqnarray}
\frac{1}{\sqrt{\snr}}\frac{d\phantom{y}}{dy}\E_X[x|y] = 
\E_X[x^2|y]-\E_X[x|y]^2\, .\label{eq:FDT}
\end{eqnarray}
Both of these formulae are obtained through simple calculus.
In order to prove Eq.~(\ref{eq:mmse}) we use (\ref{eq:twoderivatives})
in Eq.~(\ref{eq:shortentropy}) and integrate by parts with respect to $y$.
After re-ordering the various terms, we get
\begin{eqnarray}
\gexit = \frac{1}{2\sqrt{\snr}}\int\!\!\!\int \E_X[x|y]\, P(x)\, 
\frac{d\phantom{y}}{dy}Q(y|x)\; dx\; dy\, .
\end{eqnarray}
At this point we integrate by parts once more with respect to $y$ 
and use (\ref{eq:FDT}) to get the desired result.

Notice that the strikingly simple relation (\ref{eq:mmse}) was recently 
used in an iterative coding 
setting by Bhattad and Narayanan~\cite{Narayanan04}.
%
%
\itwacknowledgments

The work of A.~Montanari was partially supported by the European Union under 
the project EVERGROW.
\end{itwpaper}

\begin{itwreferences}

\bibitem{MU03b} C.~M\'easson and R.~Urbanke,
\newblock ``An upper-bound for the ML threshold of iterative coding systems 
over the BEC'',
\newblock {\em Proc. of the $41^{\text{st}}$ Allerton Conference on Communications,
Control and Computing}, Allerton House, Monticello,  USA, October 1--3, 2003.
\bibitem{Ten01} S.~ten~Brink,
\newblock ``Convergence behavior of iteratively decoded parallel concatenated codes'', \newblock {\emph{ IEEE Trans. on Communications}}, vol. 49, no. 10, October 2001.
\bibitem{LMSS01a} M.G. Luby, M. Mitzenmacher, M.A. Shokrollahi and 
D.A. Spielman,
\newblock ``Efficient erasure correcting codes'',
\newblock {\em IEEE Trans. on Information Theory},  vol. 47, no. 2,  
pp. 569--584, February 2001.
\bibitem{RSU01} T.J. Richardson, M.A. Shokrollahi  and R. Urbanke,
\newblock ``Design of capacity-approaching irregular low-density parity-check 
codes'',
\newblock {\em IEEE Trans. on Information Theory},  vol. 47, no. 2,  pp. 619--6
37, February 2001.
\bibitem{AKtB02} A.~Ashikhmin, G.~Kramer and S.~ten~Brink,
\newblock ``Code rate and the area under extrinsic information transfer curves'', \newblock {\em Proc. of the 2002 ISIT, Lausanne, Switzerland},  
June 30--July 5, 2002. 
\bibitem{Mon01} S. Franz, M. Leone, A. Montanari and F. Ricci-Tersenghi,
\newblock{``The dynamic phase transition for decoding algorithms''}, 
\newblock{Phys. Rev. E 66, 046120, 2002}.
\bibitem{Mon01b} A. Montanari,
\newblock{``Why "practical" decoding algorithms are not as good as "ideal" ones?''}, {\em DIMACS Workshop on Codes and Complexity}, Rutgers University, Piscataway, USA, December 4--7, 2001.
\bibitem{KK02} C.~Kittel and H.~Kroemer,
\newblock{{\em Thermal physics,  $2^{nd}$ Edition}}, W. H. Freeman and Co., New
 York, March, 1980.
\bibitem{Guo04} D.~Guo, S.~Shamai and S.~Verdu, 
\newblock{``Mutual information and MMSE in Gaussian channels,''}
 \newblock {\em Proc. 2004 ISIT, Chicago, IL, USA, June 2004} p. 347.
\bibitem{Narayanan04} K.~Bhattad and K.R.~Narayanan, 
\newblock{``An MSE Based Transfer Chart for Analyzing Iterative Decoding''}, 
\newblock {\em Proc. of the $42^{\text{nd}}$ A Allerton Conference on Communications,
Control and Computing}, Allerton House, Monticello,  USA, September 29--
October 3, 2004.
\end{itwreferences}

\end{document}